\documentclass{icrc2009}

\usepackage{graphicx}   
\usepackage{caption}    
\usepackage[font=footnotesize]{subfig} 
\usepackage{fixltx2e}
\usepackage{url}

\newcommand{\shorttitle}[1]%
{\markboth{Proceedings of the 31\MakeLowercase{$^{st}$} ICRC, {\L}\'{o}d\'{z} 2009}{#1} }
\newcommand{\etal}{\MakeLowercase{\textit{et al. }}} 

\begin{document}
\title{GRB Observations with the MAGIC Telescopes}

\author{\IEEEauthorblockN{	M. Garczarczyk\IEEEauthorrefmark{1},
                         				M. Gaug\IEEEauthorrefmark{2},
                            				A. Antonelli\IEEEauthorrefmark{3},
                          				D. Bastieri\IEEEauthorrefmark{4},
						J. Becerra-Gonzalez\IEEEauthorrefmark{2},\\
						S. Covino\IEEEauthorrefmark{5},
						A. La Barbera\IEEEauthorrefmark{3},
						A. Carosi\IEEEauthorrefmark{3},
						N. Galante\IEEEauthorrefmark{6},
						F. Longo\IEEEauthorrefmark{7},
						V. Scapin\IEEEauthorrefmark{8},
						S. Spiro\IEEEauthorrefmark{4}}
						for the MAGIC collaboration
						
						\\ \\
						
\IEEEauthorblockA{\IEEEauthorrefmark{1}IFAE, Edifici Cn, Campus UAB, 08193 Bellaterra, Spain}
\IEEEauthorblockA{\IEEEauthorrefmark{2}Instituto de Astrof\'isica de Canarias, via L\'actea s/n, 38205 La Laguna, Tenerife, Spain}
\IEEEauthorblockA{\IEEEauthorrefmark{3}INAF / Rome Astronomical Observatory, Via Frascati 33, 00044, Monte Porzio (Roma), Italy}
\IEEEauthorblockA{\IEEEauthorrefmark{4}Universit\`a di Padova and Istituto Nazionale di Fisica Nucleare (INFN), 35131, Padova, Italy}
\IEEEauthorblockA{\IEEEauthorrefmark{5}INAF / Brera Astronomical Observatory, Via Bianchi 46, 23807, Merate (LC), Italy}
\IEEEauthorblockA{\IEEEauthorrefmark{6}Max-Planck-Institut f\"ur Physik, F\"ohringer Ring 6, 80805 M\"unchen, Germany}
\IEEEauthorblockA{\IEEEauthorrefmark{7}Dipartimento Fisica and INFN Trieste, 34127 Trieste, Italy}
\IEEEauthorblockA{\IEEEauthorrefmark{8}Universit\`a di Udine, and INFN Trieste, 33100 Udine, Italy}
}

\shorttitle{M. Garczarczyk \etal GRB Observations with MAGIC}
\maketitle

\begin{abstract}
MAGIC is currently the most suitable instrument to perform observations of the prompt and early afterglow emission from Gamma Ray Bursts (GRBs) at energies above 25~GeV. 
The instrument is designed to have the lowest possible energy threshold and fastest reaction time to GRB Coordinates Network (GCN) alerts. 
The MAGIC-I telescope started to follow-up GCN alerts in the beginning of 2005. Since then, more than 50 GRB candidates were observed. 
Just now MAGIC-II, the second telescope has started to operate. Both telescopes together will have a factor of three better sensitivity. 
\end{abstract}

\section{VHE emission from GRBs}
GRBs are the most luminous explosions in the universe. Since their accidental discovery, now almost half a century ago, thousands of GRBs were detected 
by satellites such as BATSE, EGRET, HETE-2, Konus-Wind, INTEGRAL, Swift, AGILE and Fermi. The prompt localization by the more recent satellites and distribution 
of the coordinates over the GCN allows to observe the explosions by ground-based experiments and helped to discover the afterglow phase at lower energies. 
Theoretical models describing the afterglow phase profited from these large sample of observations at different energy bands. 

The short duration (up to few hundreds of seconds) of the prompt emission phase makes their simultaneous observation with ground based telescopes very difficult. 
The energy release during the prompt emission phase peaks in the keV-MeV regime. Since low energy emission was only observed during the decaying afterglow phase, 
the extension of the observation of the prompt emission towards the VHE regime and the measurement of cutoff energies is thus very important to allow further studies 
of the GRB origin.

Because GRBs are so powerful, they can be observed from large distances, in fact most of the GRBs have redshifts $z>1$. 
The redshift of GRB090423, for instance, was $z=8.1$~\cite{090423}, currently the most distant source ever observed up to now. 
Observations of the cutoff energy caused by the interaction of the VHE $\gamma$-rays with photons of the Extragalactic Background Light (EBL) will help 
to study the interstellar medium and its evolution with time.

Satellite detectors have large fields of view, however they are limited in size and thus in sensitivity in the VHE regime. 
The LAT detector onboard the Fermi satellite is currently the best space-born detector sensitive in the energy range between 20~MeV and 300~GeV. 
In the first eight months of operation, six GRBs were detected by LAT: GRB080825C~\cite{080825c}, GRB080916C~\cite{080916c}, GRB081024B~\cite{081024B}, 
GRB090217~\cite{090217}, GRB090323~\cite{090323} and GRB090328~\cite{090328}. All these observations showed emission extending up to few GeV. 
In the case of GRB080916C, 14 events with energies $>1$~GeV were detected~\cite{grb080916c}. In this particular burst the high energy (HE) emission 
extended up to 1400~s after the burst onset $T_{0}$, when reaching the detector sensitivity. 
The photon with the highest energy was detected at 13.2~GeV. It should be mentioned that the redshift of this burst was $z \sim 4.35$, 
for which VHE $\gamma$-rays are strongly absorbed by the EBL. All these observations indicate that VHE photons may be delayed with respect to the keV-MeV emission.

Two mechanisms can describe the measured spectra extended to HE, as observed from GRBs, so far:
\begin{enumerate}
\item In the case of GRB080916C, the HE flux follows a constant power-law decay, consistent with the Band function~\cite{band}, 
suggesting that only the synchrotron emission mechanism dominates.
\item In the case of GRB941017 and GRB970417~\cite{gonzalez} the delayed HE component shows a distinct origin 
and is best described by the Synchrotron Self Compton (SSC) mechanisms.
\end{enumerate}



\section{The performance of the MAGIC-I telescope}

The MAGIC-I telescope~\cite{magic} has a mirror area of $234 \, \mathrm{m}^{2}$ and is currently the largest Imaging Atmospheric Cherenkov Telescope (IACT) in the world. 
Its camera is equipped with photomultipliers whose photon sensitivity was enhanced with a special lacker that provides peak Quantum Efficiencies (QE) above $30\%$,
around 400~nm. The parabolic mirror of MAGIC makes parallel photons arrive synchronously at the camera and allows for very fast coincidence triggers 
that reject efficiently fake triggers from the night sky background and allow thus to reduce the trigger threshold for $\gamma$-ray showers. 
The trigger and readout system~\cite{trigger} has been optimized to fully exploit the profiles of the Cherenkov photon arrival times. 
The large mirror area, high QE photomultipliers and sophisticated trigger and readout logic provides MAGIC-I with an energy threshold well below 100~GeV.

The capabilities of MAGIC-I were demonstrated with the detection of the most distant object emitting $\gamma$-rays above 50~GeV, 
the Quasar 3C~279 at a redshift of $z=0.536$~\cite{3c279}. More recently,  a new sum trigger was implemented, which lowered the energy threshold to 25~GeV 
and lead to the first detection of the Crab Pulsar at VHE~\cite{crab}. After this proof of principle, the new sum trigger system is used during 
most GRB follow-up observations by the MAGIC-I telescope.

The lightweight design of its supporting cradle together with the optimized steering system~\cite{drive} allows MAGIC-I to slew very fast to any position in the sky. 
During the design phase of the second MAGIC-II~\cite{magic2} telescope, improvements of the drive system were made. The new steering electronics and optimization 
of the control loop parameters made the system more robust and allowed to increase the repositioning speed. Consequently, also the MAGIC-I drive system was recently upgraded 
and allows to reach the opposite location of the sky ($180^{\circ}$ movement in azimuth) now in only 20~s, while it was 54~s before the upgrade.

Compared to $\gamma$-ray satellites, IACTs have a very limited field of view (3.5$^\circ$ as in the case of MAGIC) and therefore rely on external GRB triggers, 
as the ones provided by the GCN. The GCN information is sent directly over a TCP/IP internet socket connection to the MAGIC GRB alert program. 
The program validates the alert with the predefined observability criteria. The alert is accepted and sent directly to the telescope central control if:

\begin{itemize}
\item The Sun is a zenith angle of $> 103^{\circ}$, part of the astronomical twilight is thus explicitly included in the allowed observation time.
\item The angular distance from the GRB to the Moon is $> 30^{\circ}$.
\item The zenith angle for the GRB observation is $< 60^{\circ}$. In case of moon shine the maximal zenith angle is reduced to $55^{\circ}$.
\item If the uncertainty of the GRB coordinates is larger than $1.5^\circ$, the telescope observes 15 minutes and aborts observation unless a more precise position update 
arrives.
\end{itemize}

Nominal observation durations are from the start of observability until $T_0 + 4\,\mathrm{hours}$. As the redshift of the source is normally only known 
a few hours to days later, one is obliged to observe all candidates, although a later redshift measurement can classify the observation as useless. 
As the energy threshold of an IACT depends sensitively on the observation zenith angle, most GRBs are observed with higher thresholds than the lowest possible.

\section{GRBs observed by MAGIC-I}

For the first years of operation the majority of the GRB observations by MAGIC-I were triggered by the Swift BAT detector. Since the end of 2008, 
the Fermi GBM detector started to dominate the alert frequency. Within the last observation cycle of MAGIC (May 2008 till June 2009) 
in total 251 Fermi, 155 Swift, 12 INTEGRAL and 6 AGILE GCN alerts were received by MAGIC. 
Out of these, 14 Swift, 12 GLAST and one INTEGRAL alerts fulfilled the observation criteria described above.

Due to technical tasks and bad weather not all alerts could be followed-up. 
Table~\ref{tab:magicgrbs} summarizes some of the technical details of the latest MAGIC observations~\footnote{For up-to-date information see \url{http://mojorojo.magic.iac.es/grb}}. Past observation results can be found in~\cite{grb1,grb2,grb3,grb4,grb5}.

\begin{figure}[!h]
\centering
\includegraphics[width=3in]{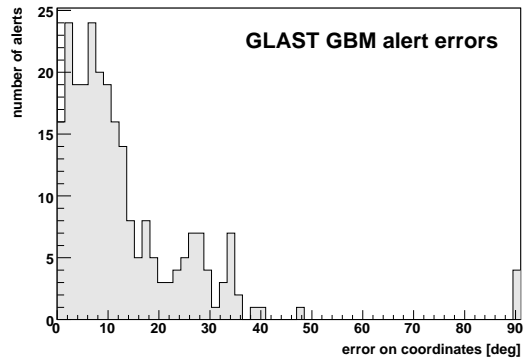}
\caption{Location accuracy of the last updated GBM alert of all 251 notices up to now. Only in $6 \%$ of the cases the alert coordinates have a precision better 
than $2^{\circ}$.}
\label{fig:GBMerrors}
\end{figure}

The GBM detector on board of the Fermi satellite provides different types of alerts. The information is updated sequently within the first tens of seconds, 
updating the coordinates and reducing their errors with the more advanced analysis. Figure~\ref{fig:GBMerrors} shows the final localization errors from all 
GBM triggers up to May 2009. Due to the large uncertainty most of the GBM alerts do not have any counterpart from follow-up observations.

MAGIC-I follows all these alerts and updates its orientation as more accurate coordinates are available. 
If the final update is not more precise than $1.5^{\circ}$, the observation is stopped automatically after 15~min, 
otherwise the burst location is observed for four hours counting from the burst onset time $T_{0}$.

The MAGIC GRB group members are assigned as burst advocates for the upcoming observations. The data is analyzed by the burst advocate within two days and, 
depending on the data quality, the result and importance of the observation, preliminary results may be published as GCN circulars.

In none of the 51 MAGIC-I GRB follow-up observations up to now, significant emission of VHE $\gamma$-rays could be detected. 
For the bursts GRB080319A~\cite{grb080319a}, GRB080330~\cite{grb080330}, GRB090102~\cite{grb090102} and GRB090113~\cite{grb090113} preliminary results were published 
as GCN circulars. These and in addition the results from GRB080315, GRB080430 and GRB080603B are summarized in table~\ref{tab:magicresults}.

GRB080315 is one of the candidates, results of which were not published as a GCN circular. The burst was triggered by BAT detector, however due to the marginal 
detection it is not clear if it was a real event. No counterparts at other wavelengths were detected. MAGIC-I started the follow-up observation 
at $T_0 + 214 \, \mathrm{s}$, short after the end of the prompt emission phase $T_{90} = 65 \, \mathrm{s}$. 
The sensitivity of the observation was reduced due to moonlight.

Two more publications are included in these proceedings. In~\cite{convino} we present the results of the GRB080430 observation. 
The MAGIC-I observation window overlapped with the afterglow observation of various experiments. Due to the well sampled data set and the relatively 
close redshift $z=0.758$ of this burst we were able to predict the afterglow emission of an SSC model, which fits the optical and X-ray data. 
Due to the late observation start (GRB080430 exploded while the Sun was still shining on the MAGIC site), our upper limits at the threshold energy 
of $E_{thr} = 80 \, \mathrm{GeV}$ are well above the predicted emission of this model.

Generally, all observations in 2009 were carried out with the new sum trigger. However, only GRB090102 had precise coordinates and a measured redshift 
such that meaningful upper limits could be obtained, with an analysis threshold energy of only $E_{thr} = 32 \, \mathrm{GeV}$. 
A more precise treatment of this burst is shown in~\cite{gaug}.

\begin{table*}[h!]
\begin{center}
\begin{tabular}{r|c|c|c|c|c|c|c|c|c|c}
     & GRB     & trigger & accuracy & $z$ & $T_{90} \, [\mathrm{s}]$ & $\Delta \mathrm{Az}$ $[\mathrm{deg}]$ & rep. time $[\mathrm{s}]$ & Zd [$\mathrm{deg}$] & start obs. $[\mathrm{s}]$ & obs. time $[\mathrm{s}]$ \\ \hline
 1. 	& 070920 	& BAT	& 0.05	& --	& 56  & -- 	& -- 		& 53 		& 550 	& 2161\\
 2. 	& 071112C 	& BAT	& 0.05	& 0.8	& 15  & 48 	& 17 		& 60 		& 4327 	& 10236\\
 3. 	& 080315 	& BAT	& 0.05 	& --	& 65  & 251 	& 69 		& 32 		& 214 	& 3262\\
 4. 	& 080319A 	& BAT	& 0.05 	& --	& 64  & 27 	& 27 		& 35	 	& 290 	& 1736\\
 5. 	& 080330 	& BAT	& 0.05	&1.51	& 61  & 129 	& -- 		& 48 		& 91 	& 1754\\
 6. 	& 080430 	& BAT	& 0.05	&0.767 	& 16  & -- 	& --		& 23 		& 4753 	& 13215 \\
 7. 	& 080603B 	& BAT	& 0.05	&2.7	& 60  & -- 	& -- 		& 41 		& 5578	& 6816 \\
 8. 	& 080605 	& BAT	& 0.05	&1.6 	& 20  & 167 	& 48 		& 37 		& 66 	& 14155 \\
 9. 	& 080903 	& BAT	& 0.05	&1.6 	& 50  & -- 	& -- 		& 55 		& 8430 	& 6135 \\
10. 	& 081028 	& BAT	& 0.05	& 3.0 	& 260 & --  	& -- 		& 60 		& 9605 	& 4711 \\
11. 	& 081130A 	& GBM	& 24.7	& -- 	& 12  & 18      & 313           & 39            & 501   & 5553 \\
12. 	& 081206A 	& GBM	& 9.6 	& -- 	& 24  & 104     & 137           & 43            & 192   & 1251 \\
13. 	& 081206C 	& GBM	& 28.5 	& -- 	& 20  & 96      &  77           & 54            &  95   & 1442 \\
14. 	& 081229  	& GBM	& 8  	& -- 	& 0.5 & 241     & 191           & 50            & 194   & 13471 \\
15. 	& 090102 	& GBM	& 4 	&1.55 	& --  & 262 	& 235 	        & 10 		& 255 	& 1161 \\
 	& 090102 	& BAT	& 0.05 	&1.55 	& 27  & --	& -- 		& 8 		& 1161 	& 13149 \\
16. 	& 090113 	& BAT	& 0.05 	& -- 	& 9.1 & 	& -- 		& 5 		& 4603 	& 9405 \\
17. 	& 090126B	& GBM	& 9.7 	& -- 	& 10.8 & 152    & 123           & 12            & 143   & 1458 \\
18. 	& 090203	& GBM	& 35.0	& -- 	& --   & 42     &  30           & 45            & 201   & 601 \\
19. 	& 090320B	& GBM	& 9.6 	& -- 	& 52  & --      & --            &               & 5010  & 480 \\
20. 	& 090323	& GBM	& 5.8	& 3.75 	& 150 & 106 	& 85 		& 20 		& 127 	& 1028 \\
21. 	& 090330	& GBM	& 90 	& --	& --  & 19 	& 17 		& 39 		& 82 	& 1230 \\
22. 	& 090401	& GBM	& 90	&-- 	& --  & -- 	& -- 		& 52 		& 108 	& 59 \\
23. 	& 090403	& GBM	& 90	&-- 	& --  & 55 	& 47 		& 43 		& 75 	& 1013 \\
\end{tabular}
\caption{Update of the GRBs observed by MAGIC. The columns from left: burst number, trigger detector, burst coordinates accuracy (detector dependent), 
redshift $z$ (if measured), burst duration $T_{90}$ (if published), repositioning distance of MAGIC in azimuth direction, required repositioning time, 
zenith distance at the beginning of the observation, begin of the MAGIC observation in respect to the burst onset $T_{0}$ and the total observation 
time with MAGIC.}\label{tab:magicgrbs}
\end{center}
\end{table*}

\begin{table*}[!h]
\centering
\begin{tabular}{|c|c|c|c|c|}
\hline
&  \multicolumn{4}{c|}{Energy bin [GeV]} \\  \cline{2-5}
\raisebox{1.5ex}[0pt]{observation time [s]}  	& 80-125 & 125-175 & 175-300 & 300-1000 \\ \hline

\multicolumn{5}{|c|}{GRB080315} \\ 
$T_0$ + 160 $\rightarrow$ $T_{0}$ + 1716	& --	& 0.16 	& 0.18 	& 0.05 \\
$T_0$ + 1761 $\rightarrow$ $T_{0}$ + 5061 	& --	& -- 	& 0.05 	& 0.05 \\ \hline

\multicolumn{5}{|c|}{GRB080319A} \\
$T_0$ + 259 $\rightarrow$ $T_{0}$ + 1736	& --	& 0.57 	& 0.11 	& 0.06 \\ \hline

\multicolumn{5}{|c|}{GRB080330} \\
$T_0$ + 91 $\rightarrow$ $T_{0}$ + 974		& --	& -- 	& 0.76 	& 0.49 \\ 
$T_0$ + 974 $\rightarrow$ $T_{0}$ + 1754	& --	& --	& --	& 0.33 \\ \hline

\multicolumn{5}{|c|}{GRB080430} \\
$T_0$ + 4753 $\rightarrow$ $T_{0}$ + 11011	& 0.55	& 0.77 	& 0.30 	& 0.05 \\ 
$T_0$ + 16912 $\rightarrow$ $T_{0}$ + 17968	& 7.76	& 6.59	& 2.45	& 0.92 \\ \hline

\multicolumn{5}{|c|}{GRB080603B} \\
$T_0$ + 5578 $\rightarrow$ $T_{0}$ + 7317	& --	& 0.16 	& 0.06 	& 0.01 \\ 
$T_0$ + 7497 $\rightarrow$ $T_{0}$ + 12357	& --	& 0.05	& 0.12	& 0.03 \\ \hline

\multicolumn{5}{|c|}{GRB090102} \\
$T_0$ + 1161 $\rightarrow$ $T_{0}$ + 5181	& 2.30	& 1.61 	& 0.34 	& 0.04 \\ 
$T_0$ + 5181 $\rightarrow$ $T_{0}$ + 9381	& 1.12	& 0.50	& 0.20	& 0.07 \\
$T_0$ + 9861 $\rightarrow$ $T_{0}$ + 11241	& --	& 1.02	& 0.30	& 0.20 \\
$T_0$ + 11241 $\rightarrow$ $T_{0}$ + 12621	& --	& 2.40	& 0.49	& 0.25 \\
$T_0$ + 12621 $\rightarrow$ $T_{0}$ + 14841	& --	& --	& 0.40	& 0.11 \\ \hline

\multicolumn{5}{|c|}{GRB090113} \\
$T_0$ + 4603 $\rightarrow$ $T_{0}$ + 10063	& 0.54	& 0.99 	& 0.10 	& 0.08 \\ 
$T_0$ + 10183 $\rightarrow$ $T_{0}$ + 13363	& --	& 0.93	& 0.23	& 0.08 \\ \hline

\end{tabular}
\caption{Upper limits from MAGIC-I GRB observations using the standard analysis approach. 
Results for different energy bins are given in units of $10^{-10} \, \mathrm{erg} \cdot \mathrm{cm}^{-2} \cdot \mathrm{s}^{-1}$.}\label{tab:magicresults}
\end{table*}

All data sets listed in table~\ref{tab:magicgrbs}, results of which are not included in table~\ref{tab:magicresults}, were rejected due to bad quality, 
either of the pointing accuracy or the weather conditions. All GBM alerts have generally too large errors to allow physical interpretation of the observation. 
In the case of GRB090323, the burst was also detected by LAT at energies reaching few GeV. The HE emission commences several seconds after the GBM trigger time, 
and lasts with some evidence for up to a couple of kilo-seconds, coinciding with the MAGIC-I observation window. 
The LAT detector is able to pinpoint the GRB sky coordinates with much higher precision. 
However, in this case the more precise coordinates were sent after the end of the MAGIC-I observation and lay outside the field of view 
of the MAGIC-I camera during the observation. 

\section{Conclusions and outlook}

The MAGIC collaboration is making many efforts to improve the experiment in order to lower the threshold energy, 
increase the sensitivity and reposition the telescope even faster for GRB follow-up observation. 
Right now the MAGIC-I telescope has undergone an upgrade of the steering system. The repositioning speed of the telescope has now reached 
the design value of 20~s for a $180^{\circ}$ movement in azimuth. The second MAGIC-II telescope is in the commissioning phase and will become operational very soon. 
Both telescopes together will have a considerably higher sensitivity at comparably low energy threshold.

\end{document}